\documentclass[runningheads,a4paper]{llncs}

\usepackage[pdftex]{graphicx}
\usepackage{amsmath}
\usepackage[caption=false,font=normalsize,labelfont=sf,textfont=sf]{subfig}
\newtheorem{assumption}{\bf{Assumption}}

\begin{document}

\title{On the Influence of Initial Qubit Placement\\During NISQ Circuit Compilation}
\titlerunning{On the Influence of Initial Qubit Placement}

\author{Alexandru Paler\inst{1}}
\institute{Linz Institute of Technology, Johannes Kepler University, Linz\\
\email{alexandru.paler@jku.at}}

\authorrunning{A. Paler}

\maketitle

\begin{abstract}
Noisy Intermediate-Scale Quantum (NISQ) machines are not fault-tolerant, operate few qubits (currently, less than hundred), but are capable of executing interesting computations. Above the quantum supremacy threshold (approx. 60 qubits), NISQ machines are expected to be more powerful than existing classical computers. One of the most stringent problems is that computations (expressed as quantum circuits) have to be adapted (compiled) to the NISQ hardware, because the hardware does not support arbitrary interactions between the qubits. This procedure introduces additional gates (e.g. SWAP gates) into the circuits while leaving the implemented computations unchanged. Each additional gate increases the failure rate of the adapted (compiled) circuits, because the hardware and the circuits are not fault-tolerant. It is reasonable to expect that the placement influences the number of additionally introduced gates. Therefore, a combinatorial problem arises: how are circuit qubits allocated (placed) initially to the hardware qubits?  The novelty of this work relies on the methodology used to investigate the influence of the initial placement. To this end, we introduce a novel heuristic and cost model to estimate the number of gates necessary to adapt a circuit to a given NISQ architecture. We implement the heuristic (source code available on github) and benchmark it using a standard compiler (e.g. from IBM Qiskit) treated as a black box. Preliminary results indicate that cost reductions of up to 10\% can be achieved for practical circuit instances on realistic NISQ architectures only by placing qubits differently than default (trivial placement).
\end{abstract}

\section{Introduction}
\label{sec:intro}

Quantum computing has become a practical reality, and the current generation of machines includes a few tens of qubits and supports only non-error corrected quantum computations. These machines, called noisy intermediate-scale quantum (NISQ), are not operating perfectly, meaning that each executed quantum gate introduces a certain error into the computed results. NISQ machines have started being used for investigations which would have been computationally very difficult previously (e.g. \cite{viyuela2018observation}). However, such investigations/computations are not fault-tolerant. It is not possible to mitigate the error through error detecting and correcting codes, because too many qubits would be required.

Executing quantum computations has become streamlined such that quantum computers are placed in the cloud and used through dedicated software tools (e.g. Qiskit and Cirq). These tools implement almost an entire work flow -- from describing quantum computations in a higher level language to executing them in a form compiled for specific architectures. Computed results are influenced by the \emph{number and type of gates} from the executed low level circuit -- output values become indistinguishable from random noise if the compiled circuits are too deep.

It is complex to compile NISQ circuits which have a minimum number of gates -- this is a classical design automation problem. Recent work on the preparation (compilation) of NISQ circuits (e.g. \cite{zulehner2018efficient}) treat this problem as a search problem. Therefore, following questions are meaningful: 1) which search criteria has to be optimised?; 2) where should the search start from?; 3) does the starting point influence the quality of the compiled circuits? Answering these questions is based on some assumptions highlighted in the following sections.

\begin{figure}[t!]
\centering
\includegraphics[width=0.5\columnwidth]{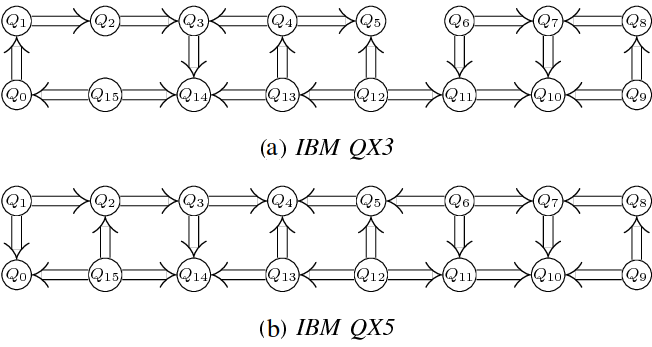}
\caption{Two architectures of IBM quantum machines where qubits are drawn as vertices and supported CNOT gates as arrows between qubit pairs: a) QX3 cannot interact through CNOT qubit pairs (5,6) and (2,15); b) QX5 has a grid like structure. In both diagrams the arrow indicate the direction of the supported CNOT: the target qubit is indicated by the tip of the arrow. The graph representation of the architectures is called \emph{coupling graph/map}.}
\label{fig:qxarch}
\end{figure}

\section{Background}
\label{sec:back}

\begin{figure*}[t!]
\centering
\subfloat[]{\includegraphics[height=0.5in]{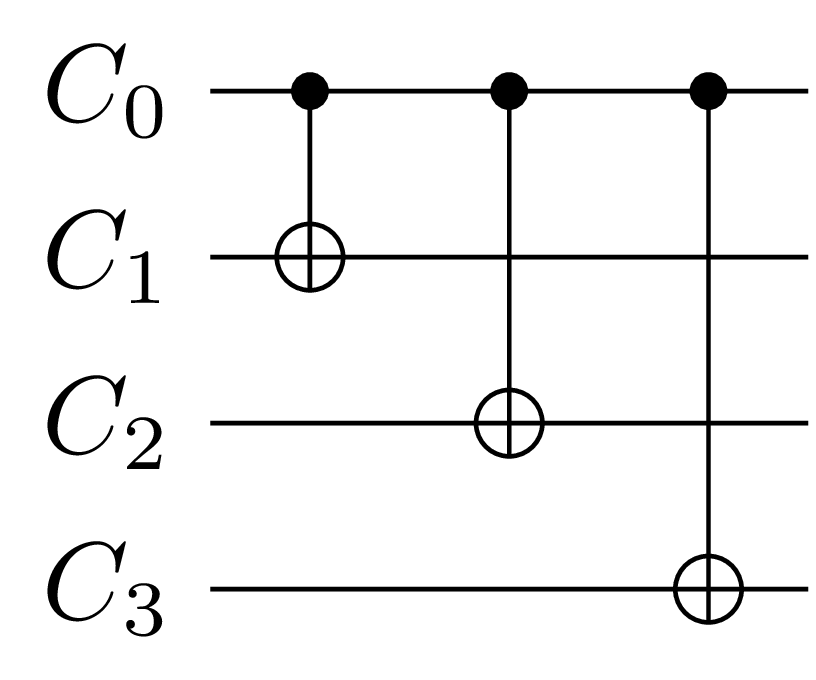}%
\label{fg_1}}
\hfil
\subfloat[]{\includegraphics[height=0.5in]{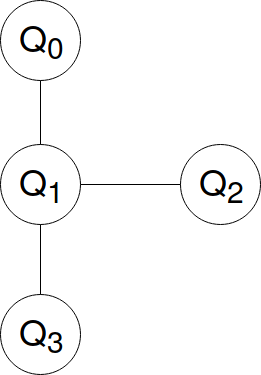}%
\label{fg_2}}
\hfil
\subfloat[]{\includegraphics[height=0.5in]{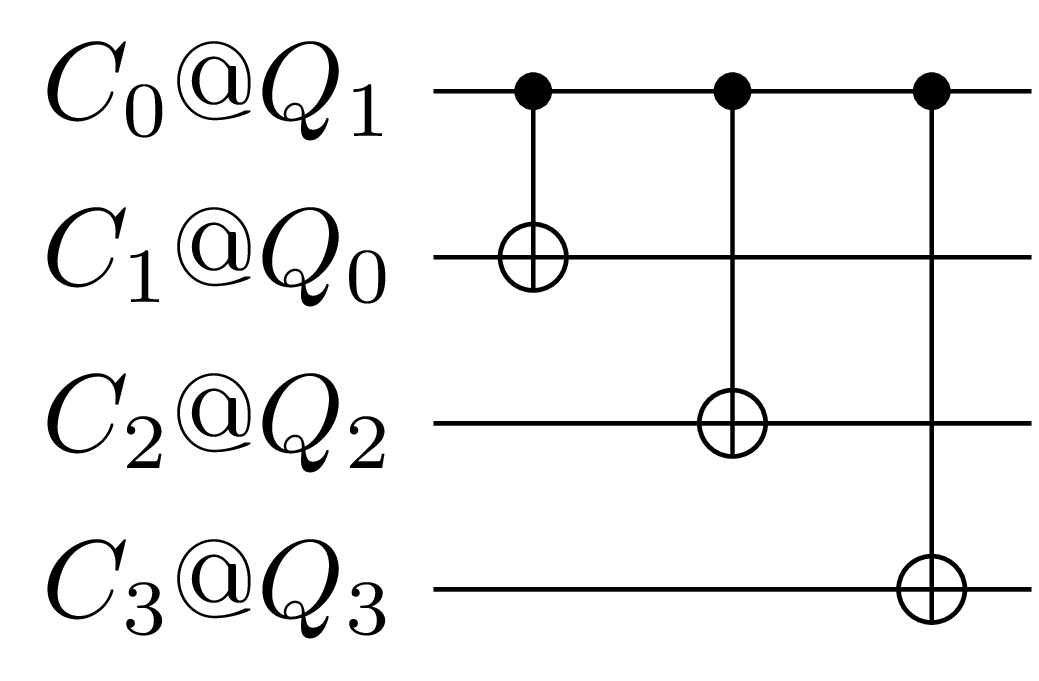}%
\label{fg_3}}
\hfil
\subfloat[]{\includegraphics[height=0.5in]{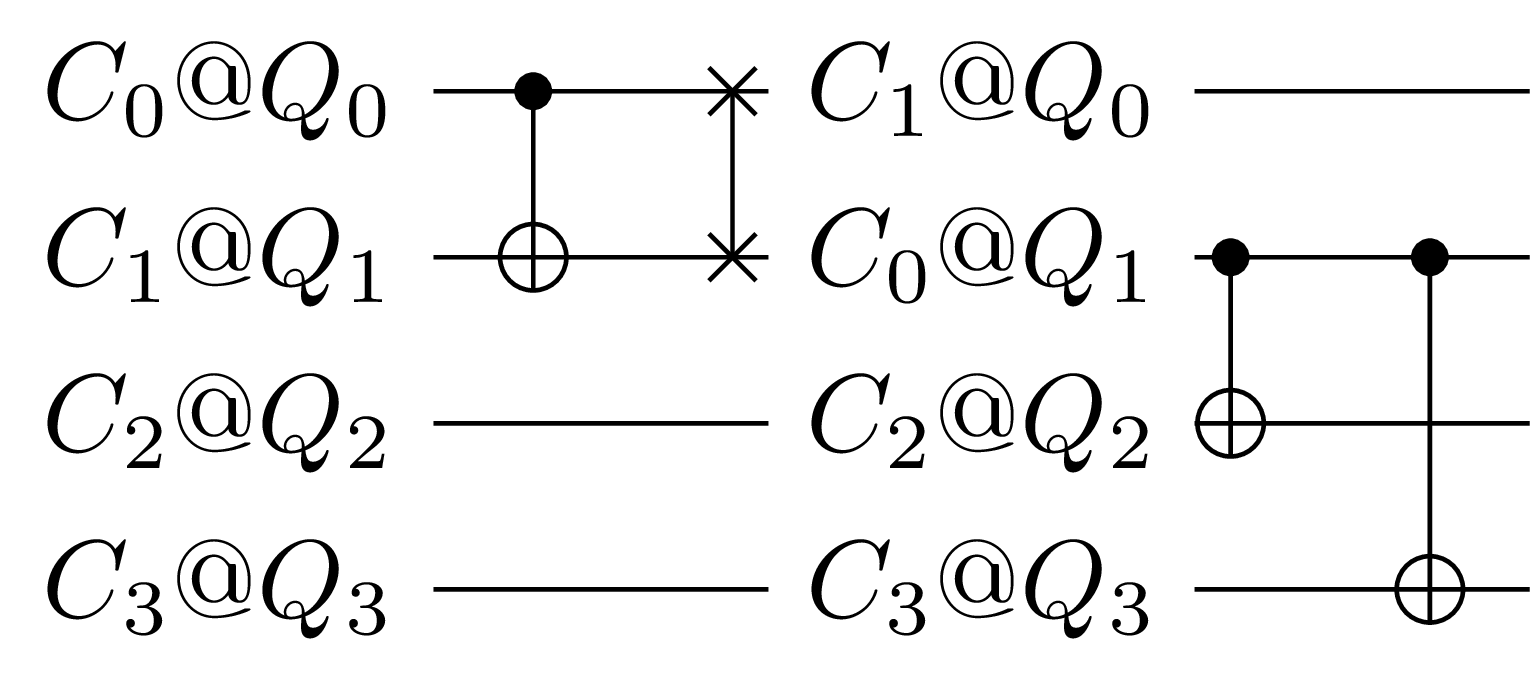}%
\label{fg_4}}
\caption{Quantum circuit examples: a) a four qubit (wire) quantum circuit; b) a four qubit coupling graph. The initial placement of wires on hardware qubits influences the number of inserted SWAP gates: c) No SWAPS inserted; d) One SWAP inserted.}
\label{fig:qcirc}
\end{figure*}

The reader is assumed to be familiar with the elements of quantum information processing and quantum circuits (wires, gates, qubits etc.). Generally, there are no restrictions how to place gates in a quantum circuit. For example, any pair of circuit wires may be used for applying a CNOT. However, NISQ machines are not flexible, and support only the application of two-qubit gates (CNOTS, in the following) between certain pairs of hardware qubits. In Fig.~\ref{fig:qxarch}, hardware qubits are drawn as vertices and the arrows represent CNOTs. Only some qubit pairs can be used for executing CNOTs (cf. QX3 with the QX5 architecture): CNOT between qubit pairs (2,15) and (5,6) are not supported by QX3. QX5 is also restrictive: for example, it does not allow a CNOT between the qubit pair (0, 13). Currently available architectures do not allow \emph{bidirectional} CNOTs (any qubit from the pair can be either the control or the target), but it is expected that this issue will not be of concern in the future. The introduced heuristic and cost model assume undirected \emph{coupling graphs} (e.g. Fig.~\ref{fg_2}).

%\subsection{Compilation}

\emph{NISQ circuit compilation} refers to the preparation of a quantum circuit to be executed on a NISQ machine expressed as a coupling graph. This procedure consists of two steps: 1) choosing an initial placement of circuit qubits to hardware qubits; 2) ensuring that the gates from the circuit can be executed on the machine. The terms machine, hardware and computer will be used interchangeably.

The \emph{qubit placement problem} is to choose which circuit wire is associated to which hardware qubit. This problem is also known as the \emph{qubit allocation} problem. For a quantum circuit with $n$ qubits (wires) to be executed on a NISQ machine with $n$ hardware qubits (vertices in the coupling graph), let the circuit qubits be $C_i$ and the machine qubits $Q_i$ with $i \in [0,n-1]$.

A \emph{trivial placement} is to map $C_0$ on $Q_0$, $C_1$ on $Q_1$ etc. One can express the placement as a vector $pl = [0, 1 \ldots, n-1]$ where wire $C_i$ is placed on hardware qubit $Q_i=pl[C_i]$. Vector $pl$ is a permutation of $n$ elements, and there are $n!$ ways how the circuit can be initially mapped to the machine. The search method (mentioned in Sec.~\ref{sec:intro}) need not use a too exact cost heuristic, because of the following assumption (a speculation).

\begin{assumption}
There are many initial placements, from all the $n!$ possibilities, which generate optimal compiled circuits.
\label{assumption:1}
\end{assumption}

After the initial placement is determined, each gate from the input circuit is mapped to the hardware. Until recently the ordering of the gates has been assumed fixed, but recent works (e.g. \cite{hattori2018quantum}) investigate how gate commutativity rules influence the compilation results. This work does not make any assumptions about how gates are mapped to the hardware. The placement of the circuit qubits dictates where SWAP gates are introduced.

\begin{example}
Consider the circuit from Fig.~\ref{fg_1} that has to be executed on the architecture from Fig.~\ref{fg_2}. If wire $C_0$ is placed on $Q_0$, the CNOT between $C_0$ and $C_2$ cannot be executed before a SWAP is inserted (Fig.~\ref{fg_4}).
\end{example}

\begin{figure}[h!]
\centering
\includegraphics[width=0.8\columnwidth]{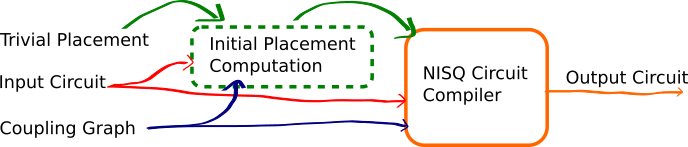}
\caption{Methodology: The compiler (orange) has three inputs and an output (the compiled circuit). One of the inputs is the initial placement of qubits (dotted green box). The potential influence of the initial placement is investigated by switching on and off the proposed heuristic.}
\label{fig:method}
\end{figure}

%exemplu cu doua configuratii care dau CNOT-uri diferite

%ideea de mismatch - care CNOT-uri nu se pot executa? - spune asta ceva despre numarul de swapuri care trebuie introduse? Este vorba despre o distanta Manhattan.

\subsection{Motivation}
\label{sec:motiv}

Qubit placement has not been considered relevant for compilation optimisation, but the significance of this problem will increase when the hardware includes many qubits. Although NISQ machines and their associated circuits are small, it is not computationally feasible to compile circuits using exact methods. This makes it difficult to determine the exact cost of an optimal solution, because finding out the exact cost means that a potential solution has already been computed. Compilation strategies will rely on heuristics and search algorithms \cite{russell2002artificial} (e.g. hill climbing, A*). As a result, the cost of compiling a circuit to NISQ should be evaluated as fast as possible and should be used as a search guiding heuristic during compilation.

\subsection{Previous Work}

Cost models for the effort of adapting a circuit to a given architecture have been intensively researched in the community (e.g. \cite{saeedi2013synthesis}), but few focused on non-exact methods and cost models (e.g. not using constraint satisfaction engines, or Boolean formula satisfiability). Most of the previous works considered regular architectures (1D or 2D nearest neighbour). 

Previous works (e.g. \cite{zulehner2018efficient} or \cite{venturelli2018compiling}) did not specifically address the initial placement problem. Nevertheless, one of the latest works where the importance of the problem is mentioned but not tackled is \cite{paler2018nisq}. Consequently, this work is, to the best of our knowledge, the first to specifically propose a solution to this problem, and to analyse its relevance in a quantitative and compiler independent manner. Independently to this work, the article of \cite{li2018tackling} was published at the same time with our paper. The difference is that we propose a different, simpler heuristic, and our methodology enables a fair interpretation of the placement influence, because the compiler is treated as a black box (Fig.~\ref{fig:method}).

\section{Methods}

This work investigates the relevance of a heuristic to approximate the number of necessary SWAPs. The three questions from Sec.~\ref{sec:intro} are reformulated into a single one: {\bf Is it possible to choose the initial placement of circuit wires on hardware qubits, such that the compiler (without being modified) generates more optimal quantum circuits?}

To answer this question we propose: a) a heuristic for the estimation of the costs, and b) a greedy search algorithm to chose the best initial placement. The output of our procedure (dotted green box in Fig.~\ref{fig:method}) is a qubit placement to be used by an unmodified NISQ circuit compiler. The default Qiskit swap\_mapper compiler was chosen, because it is a randomised algorithm. The compiler is treated as a black box with three inputs and an output (Fig.~\ref{fig:method}).

In order to simplify the presentation, the heuristic and the search method are broken into concrete steps presented in the following subsections. The proposed methodology uses all the information available: 1) input circuit $circ$, 2) coupling graph $g$, and 3) a given placement vector (a permutation) $pl$. In the following only CNOTs $cn_i$ from $circ$ are considered, and $i$ is their position index in the gate list of $circ$.

\begin{figure*}[t!]
\centering
\subfloat[]{\includegraphics[height=0.5in]{fg1.png}%
\label{gg_1}}
\hfil
\subfloat[]{\includegraphics[height=0.5in]{fg2.png}%
\label{gg_2}}
\hfil
\subfloat[]{\includegraphics[height=0.5in]{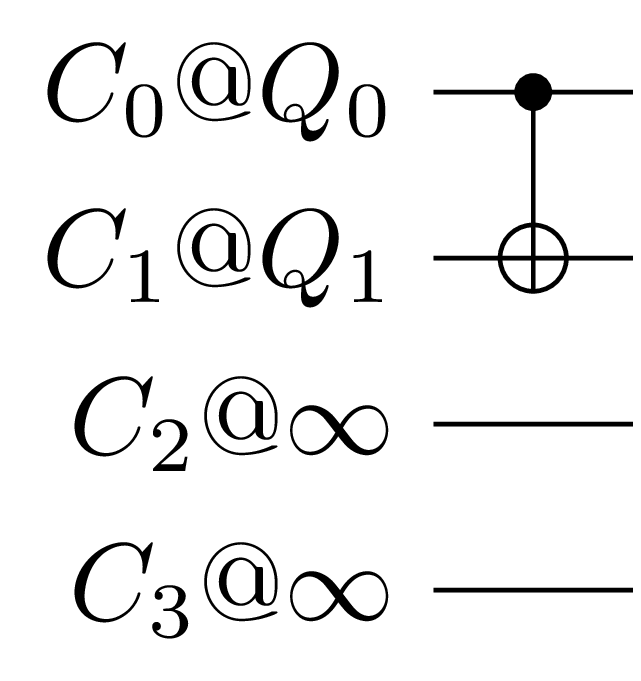}%
\label{gg_3}}
\hfil
\subfloat[]{\includegraphics[height=0.5in]{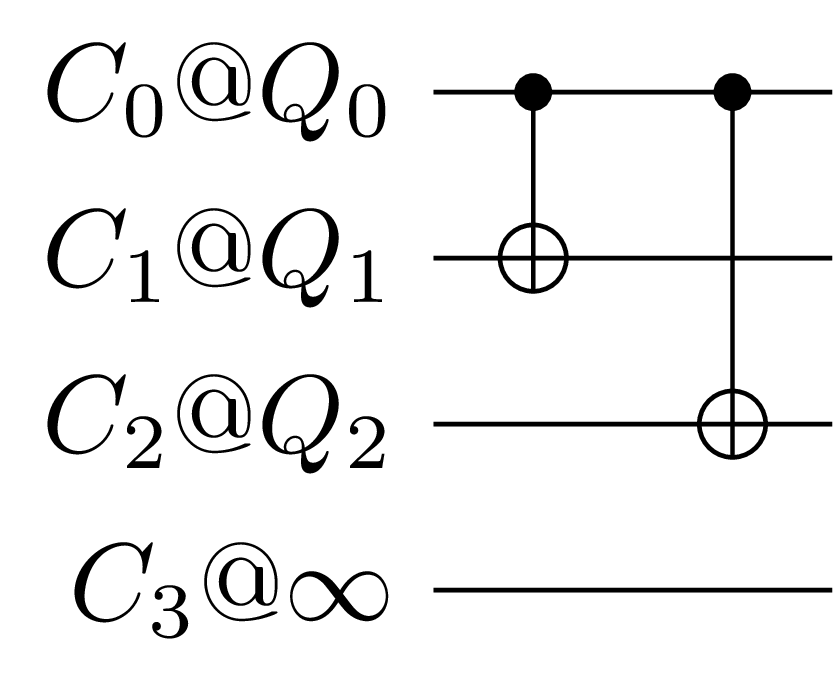}%
\label{gg_4}}
\caption{Methodology examples: a) a four qubit (wire) quantum circuit; b) a four qubit coupling graph. Sub-circuits are resulting by considering partial placements: c) $[Q_0, Q_1, \infty, \infty]$ results in a circuit with a single active CNOT; d) $[Q_0, Q_1, Q_2, \infty]$ results in a sub-circuit with two active CNOTs.}
\label{fig:qcirc2}
\end{figure*}

\subsection{Number of Active CNOTs}

\begin{assumption}
It is possible to extract a sub-circuit by removing wires and all the gates associated with the removed wires. Such a sub-circuit does not represent the original computation, but can be used to estimate costs. The minimum sub-circuit has only two wires.
\end{assumption}

For a circuit of $n$ wires, a sub-circuit is obtained using a \emph{partial placement}, where not placed wires are signaled by $\infty$. For the example of a circuit with $n=4$ wires to be mapped on an architecture with four hardware qubits, $[Q_0, \infty, \infty, Q_1]$ means that $C_0$ is placed on $Q_0$, $C_3$ on $Q_1$, while $C_1$ and $C_2$ are not placed.

For a given partial placement $pl$, the \emph{number of active CNOTs} $nr(pl)$ is the number of CNOTs operating on placed wires. In other words, this number is an indication of the number of gates $cn$ obtained from the initial circuit after removing all the wires marked by $\infty$ in $pl$.
\begin{align}
nr(circ, pl) &= |\{cn_i \text{ appl. to } C_q \text{ s.t. } pl[C_q] \neq \infty \}|
\end{align}

\begin{example}
The sub-circuits from Fig.~\ref{gg_3} and \ref{gg_4} are obtained from the initial circuit from Fig.~\ref{gg_1} by considering partial placements.
\end{example}

\subsection{Tracking the Distance}

The distances between hardware qubits is computed by a Floyd Warshall algorithm applied to $g$. For a given CNOT $cn_i$ and a coupling graph $g$, a distance function $dist(c_i, pl, g)$ computes the distance between the hardware qubits determined from $pl$. The computation method is presented instead of an expression for $dist$. The $dist$ function employs an $offset$ register of the same dimension as the number of hardware qubits.

\begin{assumption}
For a $cn_i$ having its wires placed on $Q_c$ and $Q_t$, the distance between $Q_c$ and $Q_t$ is $d_{c,t}$ which includes the Floyd Warshall distance and the values for $Q_c$ and $Q_t$ from $offset$. Both $Q_c$ and $Q_t$ have to be moved (swapped) $m=d_{c,t}/2$ times, where $m>0$. The $offset$ of $min(Q_c, Q_t)$ is updated to increase by $m-1$, and the $offset$ of $max(Q_c, Q_t)$ decreases by $m$.
\end{assumption}

The placement of the CNOT wires introduces a number of SWAP gates equal to the distance minus one. The approximation is introduced by stating that, if wire $C_w$ is placed on $pl[C_w]=Q_h$, after a sequence of CNOTs, where $Q_h$ generated $offset[Q_h]$ SWAPS, the actual hardware qubit of $C_w$ is $Q_h + offset[Q_h]$. This approximation is not ensured to be pessimistic \cite{russell2002artificial}.

\subsection{Attenuation}
\label{sec:atten}

CNOTs $cn_i$ earlier in the gate list (low values of $i$) of $circ$ should be preferred for activation instead of later CNOTs (higher values of $i$). A similar mechanism was very recently independently proposed in \cite{li2018tackling}. Attenuation is a prototypical mechanism to control the sensitivity of the cost function and a linear function was chosen. However, other functions may be more appropriate -- future work.
\begin{align}
att(cn_i, circ) &= \frac{|circ| - i}{|circ|}
\end{align}

\subsection{Approximate Cost Function}

For an input circuit $circ$, a (partial) placement $pl$, and a coupling graph $g$, the approximate cost in terms of SWAPs needed to compile it to $g$ is given by:
\begin{align}
err(circ, pl, g) &= \sum_{i=0}^{|circ|} att(cn_i, circ) \times dist(cn_i, pl, g)\\
cost(circ, pl, g) &= \frac{err(circ, pl, g)}{nr(circ, pl)}
\end{align}

As a result, for equal $err$ values, the cost is lower for partial placements that activate more CNOTs from the initial circuit. Simultaneously, for equal number of active CNOTs, lower total errors are preferred.

\subsection{Greedy Search - One Hardware Qubit After the Other}

The search procedure for finding an initial placement is iterative: it places a circuit wire on each $Q_i$ at a time. The search starts with $Q_0$, and a best wire candidate for this hardware qubit. For simplicity, assume the wire is $C_0$. Afterwards, the search starts to construct partial placements that minimise the $cost(circ, pl, g)$ function. For example, assume the best wire from $circ$ is $C_4$, because it increases the number of activated CNOTs in the sub-circuit generated by the partial placement $pl$ while keeping the error low. Thus, $C_4$ is placed on $Q_1$ (hardware qubit index following $0$ is $1$) s.t. $pl[C_4]=Q_1$.

The implemented search method generates a search tree. The root is the wire placed on $Q_0$, and the leaves are the wires placed on $Q_n$. Choosing the initial placement with the minimal cost is similar to executing a breadth first search, and this could lead to an exponential complexity. Nevertheless, the search algorithm was designed to limit the number of children (maximum number of candidates with equal estimated costs) and supports also cut-off heuristics (chose the best partial placement, when the search tree has a certain number of levels). The search heuristic is implemented modularly (can be entirely replaced) in our Python implementation.

\subsection{Discussion: Heuristic vs. Exact Solution}
\label{sec:versus}

It can be argued that the herein presented cost model and heuristic to estimate the costs can be replaced by exact methods (e.g. SAT solvers). This is true for circuits with few qubits, and this could allow an improved quantitative analysis of the proposed methodology. 

However, we argue that small circuits do not require design automation methods, because these can be designed manually (an approach taken by Cirq \footnote{github.com/quantumlib/Cirq}). NISQ platforms are very resource restricted and it is faster to design circuits by hand, than to encode all the constraints into a SAT instance. According to Assumption~\ref{assumption:1}, it is also expected (based on previous experience in designing quantum circuits) that exact methods will be slower and error prone (due to software bugs), such that software verification tools are necessary -- or manual verification of the compiled circuits is required, which again is an argument against exact methods.

For experimentation using cloud based services (e.g. IBM \footnote{quantumexperience.ng.bluemix.net/qx}) it is important to compile as well as possible, but \emph{compilation speed plays an important role}, too. Therefore, exact methods are not scalable (fast enough) and their practicality is restricted to quantitative comparisons on a set of benchmark circuits. At this point it should be noted that the research community has not agreed on a set of NISQ benchmarking circuits. The circuits we used in this paper were chosen only because they have a practical relation to existing IBM architectures.

It is impractical to manually design large NISQ circuits (e.g. more than 100 qubits), and at the same time it is not possible (due to scalability limitations) to use exact methods for such circuits.

Consequently, we argue that exact solutions are theoretically useful, but impractical. Heuristics are the only practical and scalable solution for the compilation and optimisation of NISQ circuits. Therefore, \emph{heuristics can be compared only against heuristics}. This work, besides \cite{li2018tackling}, is one of the very few to tackle qubit placement from a heuristics point of view.

\section{Results}
\label{sec:results}

The proposed heuristic for choosing the initial placement was implemented in Python and integrated into the Qiskit SDK challenge framework\footnote{https://qe-awards.mybluemix.net/static/challenge\_24\_1\_2018.zip}. Simulation results were obtained for four circuits: two random circuits (rand0\_n16\_d16 and rand1\_n16\_d16) of 16 qubits, and two practically relevant circuits (qft\_n16 and qubits\_n16\_exc). Each of the four circuits was compiled for both the IBM QX3 and QX5 architectures. Each compilation was performed twice: the first time with the trivial initial placement, and the second time with a placement chosen by the presented heuristic. Two variants of the heuristics were used: one with attenuation and the second without. A total of 250 simulations were performed for each combination of circuit, architecture, and placement. Each combination used the same seed -- for example, simulation number $15$ used a seed value of $15$.

\begin{table}[!t]
\scriptsize
\caption{Simulation results with attenuation (\ref{sec:atten}). The \emph{T} column is the trivial placement cost, and \emph{H} the initial placement computed by the heuristic. The \emph{Med.} columns include the median value from 250 simulations, and the the \emph{Avg.} columns the average value. \emph{Imp.} is computed as the ratio between \emph{T.Avg.} and \emph{H.Avg.}.}
\label{tbl:wi}
\centering
\begin{tabular}{l|l|r|r|r|r|r}
\hline
Circuit & Arch. & T.Avg.& T.Med. & H.Avg. & H.Med. & Imp.\\
\hline
qft\_n16 & QX3 & 7837.76 & 7811.00 & 7469.70 & 7387.00 & 1.05\\
qft\_n16 & QX5 & 7142.37 & 7092.00 & 7265.49 & 7178.00 & 0.98\\
qubits\_n16\_exc & QX3 & 1447.22 & 1422.00 & 1541.15 & 1522.00 & 0.94\\
qubits\_n16\_exc & QX5 & 562.00 & 562.00 & 562.00 & 562.00 & 1.00\\
rand0\_n16\_d16 & QX3 & 13094.18 & 13078.00 & 13492.68 & 13420.00 & 0.97\\
rand0\_n16\_d16 & QX5 & 12370.19 & 12326.00 & 12243.39 & 12261.00 & 1.01\\
rand1\_n16\_d16 & QX3 & 14188.84 & 14209.00 & 13604.65 & 13569.00 & 1.04\\
rand1\_n16\_d16 & QX5 &12779.25 & 12810.00 & 12542.53 & 12537.00 & 1.02
\end{tabular}
\end{table}

\begin{table}[!t]
\scriptsize
\caption{Simulation results without attenuation. The \emph{T} column is the trivial placement cost, and \emph{H} the initial placement computed by the heuristic. The \emph{Med.} columns include the median value from 250 simulations, and the the \emph{Avg.} columns the average value. \emph{Imp.} is computed as the ratio between \emph{T.Avg.} and \emph{H.Avg.}.}
\label{tbl:wo}
\centering
\begin{tabular}{l|l|r|r|r|r|r}
\hline
Circuit & Arch. & T.Avg.& T.Med. & H.Avg. & H.Med. & Imp.\\
\hline
qft\_n16 & QX3 & 7837.76 & 7811.00 & 7205.67 & 7137.00 & 1.09\\
qft\_n16 & QX5 & 7142.37 & 7092.00 & 6859.87 & 6867.00 & 1.04\\
qubits\_n16\_exc & QX3 & 1447.22 & 1422.00 & 1763.40 & 1773.00 & 0.82\\
qubits\_n16\_exc & QX5 & 562.00 & 562.00 & 1050.51 & 1020.00 & 0.53\\
rand0\_n16\_d16 & QX3 & 13094.18 & 13078.00 & 13013.54 & 12995.00 & 1.01\\
rand0\_n16\_d16 & QX5 &12370.19 & 12326.00 & 12085.82 & 12080.00 & 1.02\\
rand1\_n16\_d16 & QX3 & 14188.84 & 14209.00 & 12851.42 & 12879.00 & 1.10\\
rand1\_n16\_d16 & QX5 & 12779.25 & 12810.00 & 12121.74 & 12175.00 & 1.05
\end{tabular}
\end{table}

The results validate to some degree the predictive power of the approximative cost model (see previous Section), and indicates the sensitivity of future compilation search algorithms to the initial placement of the circuit qubits.

The costs from the Tables~\ref{tbl:wi} and \ref{tbl:wo} are the sum of all specific gate type costs of each output circuit gate. The CNOTs have a cost of 10, single qubit gates a cost of 1 (the exception is formed by single qubit Z-axis rotations which have cost zero). The SWAP gates are decomposed into three CNOTs and four Hadamard gates. The columns with \emph{T} refer to the trivial placement, and \emph{H} to the initial placement computed by the heuristic. The \emph{Med.} columns include the median value from the 250 simulations, and the the \emph{Avg.} columns the average value. The \emph{Imp.} columns are computed as the ratio between the values from the corresponding \emph{T.Avg.} and \emph{H.Avg.}. For example, an improvement of $1.05$ means that the heuristic reduced cost by 5\% (positive influence), while $0.55$ means that the costs increased by 45\% (negative influence).

Across all experiments, the median and the average values are almost equal, suggesting that there are no outliers. Thus, the observed improvements (positive and negative) seem correlated with the usage of the heuristic. Some preliminary conclusions are: 1) for practical circuits, the heuristic without attenuation generates diametrically opposed results (10\% and 4\% improvements for QFT and -18\% and -47\% increases for the other circuit) -- this is \emph{a good result in the case of QFT, because it is a widely used sub-routine in quantum algorithms}; 2) \emph{the proposed heuristic seems to work better for architectures which do not have a completely regular structure} (cf. QX3 to QX5) -- this is a good feature, because such architectures are the complex situations; 3) on average, the heuristic does not influence the overall performance of the compiler, but \emph{for specific circuit types it has significant influence} (while some are improved by up to 10\%, others are worsened by 45\%) -- the proposed heuristic is sensitive to a structural property of the circuits which needs to be determined.

The qubits\_n16\_exc circuit is a kind of black sheep of the experimented circuits: the proposed algorithm does not perform better than the defaults. Preliminary investigations show that this is happening because the circuit has a very regular structure, it actually uses only 14 qubits from 16 (leaving two un-operated, although defined in the quantum registers of the circuit), and the ordering of the qubits is (by chance?) already optimal for the considered architectures. Thus, incidentally, the default layout is already optimal and all the heuristic efforts are worsening the qubit allocation. Future work will look at different architectures.

\section{Conclusion}
\label{sec:concl}

The results are promising when comparing them with the ones obtained by the winner of the IBM Qiskit competition\footnote{www.ibm.com/blogs/research/2018/08/winners-qiskit-developer-challenge}: a specialised compiler using advanced techniques achieved on average 20\% improvements. The proposed initial placement method generates improvements between 2\% and 10\% with a sub-optimal compiler, which IBM intended to replace by organising a competition. Our source code is available at github.com/alexandrupaler/k7m.

We did not perform a comparison against the IBM challenge winner, because it would not have been a fair one. The challenge winner uses information about the internals of the circuits (two qubit arbitrary gates are decomposed with the KAK algorithm \cite{zulehner2018compiling}) in order to partially recompile subcircuits. The herein proposed cost model and search method are agnostic to quantum circuit internals or other previous circuit decomposition techniques.

Future work will focus on comparing future heuristics that will be proposed in the literature. We will first investigate why the structure of the qubits\_n16\_exc circuit makes the heuristic perform in a suboptimal manner (cf. discussion about benchmark circuits in Sec.~\ref{sec:versus}). Second, different attenuation factor formulas will be investigated and compared to the very recent work of \cite{li2018tackling}. Finally, and most important, the effect of the heuristic will be analysed when used from within a compiler -- do such approximate cost models improve the speed and the quality of NISQ circuit compilation? The current preliminary results presented in this work suggest this could be the case. It may be necessary to improve the distance estimation and the attenuation formulas.

\section*{Acknowledgment}
\small
The author thanks Ali Javadi Abhari for suggesting some of the circuits, and Lucian Mircea Sasu for very helpful discussions. This work was supported by project CHARON funded by Linz Institute of Technology.

\bibliographystyle{IEEEtran}
\bibliography{influence}

\end{document}